\documentclass[journal]{IEEEtran}
\IEEEoverridecommandlockouts
\usepackage{cite}
\usepackage{amsmath,amssymb,amsfonts}
\usepackage{bbm}
\usepackage{algorithmic}
\usepackage{graphicx}
\usepackage[]{hyperref} 
\usepackage{textcomp}
\usepackage{url}
\usepackage{mathrsfs}
\usepackage{algorithm}
\usepackage{algorithmic}
\usepackage{multirow}
\usepackage{multicol}
\usepackage{booktabs}
\usepackage[numbers,sort&compress]{natbib}
\usepackage{siunitx}
\usepackage{booktabs, array}
\usepackage{stfloats}
\def\BibTeX{{\rm B\kern-.05em{\sc i\kern-.025em b}\kern-.08em
    T\kern-.1667em\lower.7ex\hbox{E}\kern-.125emX}}
\begin{document}

\title{Block-Wise Mixed-Precision Quantization: Enabling High Efficiency for Practical ReRAM-based DNN Accelerators
}
\author{Xueying Wu, Edward Hanson, Nansu Wang, Qilin Zheng, {\em Student Member, IEEE}, Xiaoxuan Yang, {\em Student Member, IEEE}, Huanrui Yang, {\em Member, IEEE}, Shiyu Li, {\em Student Member, IEEE}, Feng Cheng, Partha Pratim Pande, {\em Fellow, IEEE}, Janardhan Rao Doppa, {\em Senior Member, IEEE}, Krishnendu Chakrabarty, {\em Fellow, IEEE}, and Hai (Helen) Li, {\em Fellow, IEEE}.
\thanks{ This work was supported in
part by the U.S. National Science Foundation (NSF) under Grant CNS-1955196, Grant CNS-2233808, Grant EECS-2023752, Grant CNS-1822085, and Grant CSR-1955353.
Xueying Wu, Edward Hanson, Nansu Wang, Qilin Zheng, Xiaoxuan Yang, Shiyu Li, Feng Cheng and Hai Li
are with the Department of Electrical and Computer Engineering, Duke
University, Durham, NC 27708 USA (e-mail: xw221@duke.edu).
Huanrui Yang is with the EECS department of UC Berkeley, Berkeley, CA 94720 USA.
Krishnendu Chakrabarty is with the School of Electrical, Computer and Energy Engineering, Arizona State University, Tempe, AZ 85281 USA.
Janardhan Rao Doppa and Partha Pratim Pande are with the School of
Electrical Engineering and Computer Science, Washington State University,
Pullman, WA 99163 USA}}

\maketitle

\begin{abstract}
Resistive random access memory (ReRAM)-based processing-in-memory (PIM) architectures have demonstrated great potential to accelerate Deep Neural Network (DNN) training/inference. However, the computational accuracy of analog PIM is compromised due to the non-idealities, such as the conductance variation of ReRAM cells. The impact of these non-idealities worsens as the number of concurrently activated wordlines and bitlines increases. To guarantee computational accuracy, only a limited number of wordlines and bitlines of the crossbar array can be turned on concurrently, significantly reducing the achievable parallelism of the architecture.

While the constraints on parallelism limit the efficiency of the accelerators, they also provide a new opportunity for fine-grained mixed-precision quantization.
To enable efficient DNN inference on practical ReRAM-based accelerators, we propose an algorithm-architecture co-design framework called \underline{B}lock-\underline{W}ise mixed-precision \underline{Q}uantization (BWQ). 
At the algorithm level, BWQ-A introduces a mixed-precision quantization scheme at the block level, which achieves a high weight and activation compression ratio with negligible accuracy degradation. We also present the hardware architecture design BWQ-H, which leverages the low-bit-width models achieved by BWQ-A to perform high-efficiency DNN inference on ReRAM devices.
BWQ-H also adopts a novel precision-aware weight mapping method to increase the ReRAM crossbar's throughput. 
Our evaluation demonstrates the effectiveness of BWQ, which achieves a $6.08 \times$ speedup and a $17.47 \times$ energy saving on average compared to existing ReRAM-based architectures.
\end{abstract}


\section{Introduction} \label{intro}

\begin{figure}
    \centering
    \includegraphics[width=0.48\textwidth]{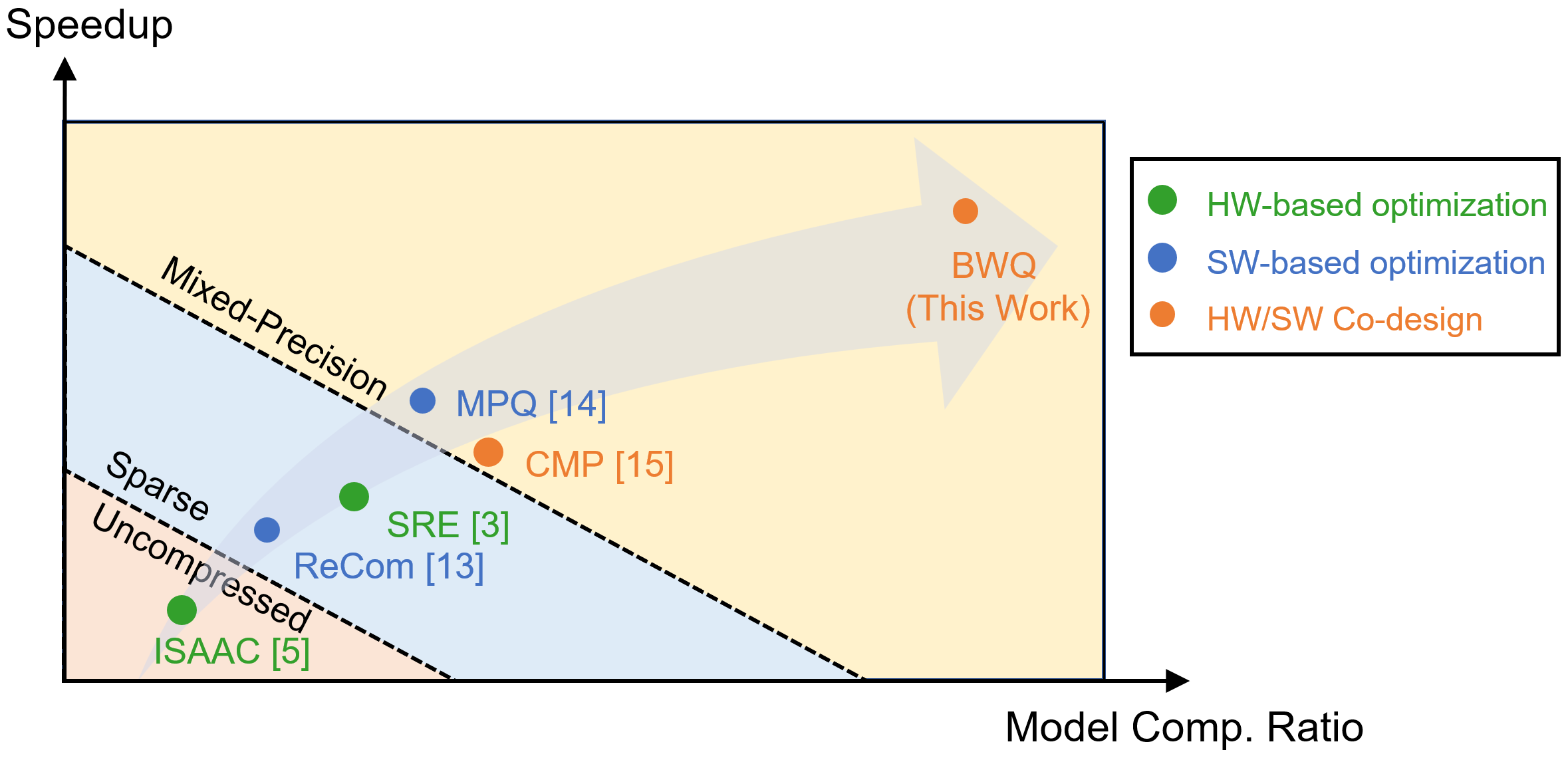}
    \caption{Research about ReRAM-based DNN accelerators from different optimization perspectives.}
    \label{intro-fig}
\end{figure}

\IEEEPARstart{R}{esistive} random access memory (ReRAM)-based Processing-in-memory (PIM) architectures can perform in-situ computation within the memory devices, and they have demonstrated great potential in accelerating DNN training/inference \cite{song2017pipelayer, yang2020retransformer}. 
However, the manufacturing technology of ReRAM devices is still in its early stage, and there exist many challenges to its practical adoption \cite{yang2019sparse}. 
Most of the works about ReRAM-based DNN accelerators have overlooked practical considerations and rely on an idealized assumption regarding ReRAM devices and associated analog-to-digital converter (ADC) overhead. 
They assume that it is possible to activate all the rows and columns of a $128\times 128$ or $256\times 256$ array simultaneously within a single clock cycle without impacting computational accuracy\cite{chi2016prime, shafiee2016isaac}. However, there are several challenges that render this assumption impractical.
The major problem is the conductance variation of ReRAM devices. Since ReRAM crossbar arrays leverage Kirchhoff’s Current Law to perform vector-matrix multiplication (VMM) operations, the conductance variation accumulated along the bitlines (BLs) is proportional to the number of concurrently activated wordlines (WLs) \cite{tu2022sdp}. Activating too many WLs simultaneously also leads to high BL current, which would induce significant IR-drop and cause non-uniform voltage and current distribution along the crossbar \cite{zheng2020mobilatice}.
Therefore, to achieve high-accuracy computation, the number of WLs that can be activated within a crossbar array simultaneously should be limited. 
Another challenge is that for practical ReRAM-based DNN accelerators, the number of ADCs for each crossbar array should be restricted as they consume a significant amount of power and area \cite{shafiee2016isaac, 10244258}. As such, it is necessary to share one ADC among multiple BLs. 
Given that an ADC can only convert the signals of one BL in a single clock cycle, the number of BLs that can be activated simultaneously should match the number of ADCs in each crossbar \cite{yang2019sparse}.
For a practical ReRAM-based DNN accelerator, the VMM on the crossbar arrays should operate at a much finer granularity, termed as an Operation Unit (OU), rather than at the subarray granularity. \cite{lin2018dl, yang2019sparse, yang2021auto}. 
It is demonstrated by several recent studies that for a practical ReRAM-based DNN accelerator to attain an acceptable level of inference accuracy, only nine WLs and eight BLs can be turned on concurrently \cite{chen201865nm, xue202015, yang2019sparse}.

The above constraints impose significant limitations on the achievable parallelism of ReRAM-based accelerators, as multiple cycles are required to finish the computation with the entire crossbar array. 
In the OU-based operation scheme, ADC latency dominates the runtime, and ADC contributes mostly to the overall energy consumption due to the increased number of cycles for computation.
Therefore, model compression methods are desired to reduce the number of computation cycles and improve the efficiency of the accelerators. 
Fig. \ref{intro-fig} provides a qualitative illustration of the performance of various ReRAM-based DNN accelerators. 
The speedup of previous studies is scaled considering the effects of the OU-based operation scheme. 
Basically, these studies can be categorized into three classes according to their optimization perspectives: hardware (HW)-based optimization, software (SW)-based optimization, and HW/SW co-design. 
HW-based optimization solutions only improve the accelerators' performance from an HW perspective, featuring intra-layer pipeline (ISAAC \cite{shafiee2016isaac}) or leveraging the natural sparsity of the neural networks with index reordering (SRE \cite{yang2019sparse}). 
SW-based optimization solution ReCom \cite{ji2018recom} and MPQ \cite{huang2021mixed} proactively compress the models with algorithms. 
However, these works fail to take into account the practical constraints of the underlying ReRAM devices and thus are not able to achieve the optimal model compression ratio.
HW/SW co-design-based solutions perform optimizations from both HW and SW perspectives and are supposed to achieve the highest performance.
However, existing solutions only compress the model with coarse-grained schemes. 
For example, CMP \cite{zhu2019configurable} conducts layer-wise mixed-precision quantization, and its model compression performance is limited.
In contrast, our work proposes a block-wise mixed-precision quantization algorithm (BWQ-A) that considers the parallelism constraints of ReRAM-based accelerators.
BWQ-A performs mixed-precision quantization at the weight block (WB) level, matching the size of the OU. 
The WBs are assigned varying bit precisions based on their individual significance, and the precisions are learned throughout the training process.  
Leveraging a much finer quantization granularity, BWQ-A can achieve a 58.27$\times$ weight compression and a 9.47$\times$ activation compression on average over the floating-point baseline models with less than 1\% of accuracy degradation.
At the architecture level, we present a ReRAM-based hardware accelerator BWQ-H, which enables the efficient inference of the models quantized using BWQ-A. 
Experimental results show that BWQ-H can boost the energy efficiency by $17.47 \times$ and reduce the latency by $6.08 \times$ on average compared to existing ReRAM-based architectures.

The main contributions of our work include:
\begin{itemize}
    \item We introduce BWQ-A, a novel mixed-precision quantization algorithm at the weight block level. Leveraging the small quantization granularity, we are able to achieve a higher weight and activation compression ratio under similar accuracy compared to the existing quantization schemes.
    \item We propose a practical ReRAM-based accelerator with the OU-based operation scheme. 
    BWQ-H enables efficient inference of the models quantized using BWQ-A. 
     It employs a novel precision-aware weight mapping scheme that increases memory utilization within an OU for the mixed-precision quantized models. A memory controller is also designed to facilitate mixed-precision computation within the same crossbar.
    \item We analyze the scalability of BWQ-A and BWQ-H with larger OU sizes. This serves as a future road map for practical ReRAM accelerators depending on the evolution of the manufacturing technology of ReRAM devices. 
\end{itemize}

The remainder of this paper is organized as follows:
In Section \ref{background}, we discuss the background of ReRAM-based DNN accelerators and review the mixed-precision quantization schemes for model compression. 
Section \ref{BWQ-A} introduces BWQ-A, the block-wise mixed-precision quantization algorithm. 
Next, Section \ref{BWQ-H} presents the architecture of the hardware accelerator BWQ-H that enables efficient inference of the BWQ-A algorithm. 
Section \ref{exp-setup} shows the evaluation methodology, and Section \ref{results} shows the performance of the algorithm and the architecture. 
Lastly, Section \ref{conclusion} concludes the paper and outlines key points for future works.

\section{Background} \label{background}
\subsection{ReRAM-based DNN Accelerators}
The potential of ReRAM-based platforms to achieve high performance and energy efficiency computation has made them promising for accelerating DNN inference.
ReRAM is a type of non-volatile memory that stores information by changing the resistance of the metal oxide material \cite{chi2016prime}. 
In ReRAM-based accelerators, the conductance of ReRAM devices is used to represent the weights of neural networks, and the analog VMM computations are performed inside the crossbar arrays. This allows for highly parallel computation and eliminates the need for data movement between memory and computation units. 


Most of the studies on ReRAM-based DNN accelerators have utilized crossbars with sizes of $128\times128$ or $256\times256$ \cite{shafiee2016isaac, chi2016prime}. These design choices strike a balance between the throughput and utilization of the ReRAM crossbars. 
These studies usually assume that it is possible to activate all the rows and columns of the array simultaneously, and the VMM computation with the whole weight matrix on the subarray could finish within a single clock cycle. 
However, this assumption is ideal. According to the experimental results in \cite{yang2019sparse}, activating the entire ReRAM crossbar within one cycle leads to significant accuracy loss and introduces high peripheral circuitry overhead.
To guarantee high-accuracy computation, it is necessary to reduce the accumulated conductance variation of the ReRAM devices and IR-drop along the BLs. Thus, the number of concurrently activated WLs should be limited. 
Additionally, the number of BLs to be turned on concurrently should also be limited to constrain the overhead of the ADCs. According to \cite{10244258}, the ADCs account for 50\%~70\% of the overall power consumption in a ReRAM-based accelerator.
For energy and area efficient implementation, one ADC should be shared among multiple BLs and the number of concurrently activated BLs should also be restricted by the number of ADCs. 
A practical ReRAM accelerator should operate at the granularity of an OU, which is much smaller than the entire crossbar. Several studies have demonstrated that, in practice, an OU can accommodate a block with only nine WLs and eight BLs \cite{chen201865nm, yang2019sparse}.



\subsection{Mixed-Precision Quantization}\label{mpq}
Quantization is a model size reduction technique that converts the floating-point weights into low-precision floating-point or integer formats.
Low precision quantization is particularly beneficial for ReRAM-based accelerators as it efficiently reduces the number of columns required to represent each weight, thus reducing the computation cycles and ADC's power consumption.
Compared with uniform quantization, mixed-precision quantization can achieve lower average bit precision under similar accuracy levels \cite{dong2019hawq, wang2019haq, wu2018mixed}. 
However, determining the optimal precision for each layer or each channel is a challenging task.
Most previous works have either performed manual bit-with selection, which relies on expert knowledge \cite{dong2019hawq}, or neural architecture search with reinforcement learning, which requires massive computation \cite{wang2019haq, wu2018mixed}. 

To address the above challenges, Bit-level sparsity quantization (BSQ) \cite{yang2021bsq} proposes a layer-wise mixed-precision quantization scheme that learns the precision of the weights in each layer through a single-pass training process. 
To achieve an optimal trade-off between model size and accuracy, BSQ proposes exploiting bit-level sparsity by training bit representation of the weights instead of floating-point weights. 
During training, a bit-level group Lasso regularizer is also incorporated to induce higher sparsity. 
As evaluated on a diverse range of models and datasets, BSQ is able to achieve higher compression ratios under similar accuracy compared to previous methods \cite{yang2021bsq}.


\subsection{Motivation}
A practical ReRAM-based accelerator's achievable throughput and energy efficiency are limited by the OU size due to the computation accuracy and the peripheral circuitry overhead concerns. The VMM with the entire weight matrix on a single crossbar array should require multiple ADC cycles as each cycle can only activate a $9\times8$ OU, which is the maximum size that a state-of-the-art ReRAM accelerator can achieve \cite{chen201865nm}.
BSQ's potential to achieve ultra-low weight precision provides a possible solution for efficient DNN inference on ReRAM-based accelerators, as the low-bit-width models greatly reduce the required ADC cycles and ADC energy consumption. 
However, BSQ only considers quantization at the layer level. 
According to Dash et. al. \cite{dash2021robust}, the significance, or the contribution towards the training objective of each weight element within the same layer can vary considerably. 
BSQ assigns a uniform bit precision to all the weight elements in a layer, which may not lead to the optimal trade-off between accuracy and average bit-width per weight.
On the other hand, the constraints for a practical ReRAM accelerator provide a new opportunity for finer-grained mixed-precision quantization. This motivates us to explore a higher weight compression ratio with a new quantization scheme, block-wise mixed-precision quantization algorithm, or BWQ-A. 
We also propose a corresponding hardware design BWQ-H to enable efficient implementation of the BWQ-A algorithm on ReRAM-based architecture.

\section{BWQ-A: Block-Wise Mixed-Precision Quantization Algorithm} \label{BWQ-A}
BSQ's potential to achieve ultra-low weight precision provides a possible solution for efficient DNN inference on ReRAM-based accelerators under the OU-based operation scheme, as the low-bit-width models effectively reduce the required clock cycles and ADC energy consumption. 
However, BSQ only performs layer-wise mixed-precision quantization, overlooking the fact that the significance, or the contribution towards the training objective of each weight element within the same layer can vary considerably \cite{dash2021robust}. 
On the other hand, the OU-based operation scheme provides a new opportunity for finer-grained mixed-precision quantization. This motivates us to explore a higher weight compression ratio with a novel quantization scheme BWQ-A.

\subsection{Weight Compression}
\begin{figure}
    \centering
    \includegraphics[width=0.48\textwidth]{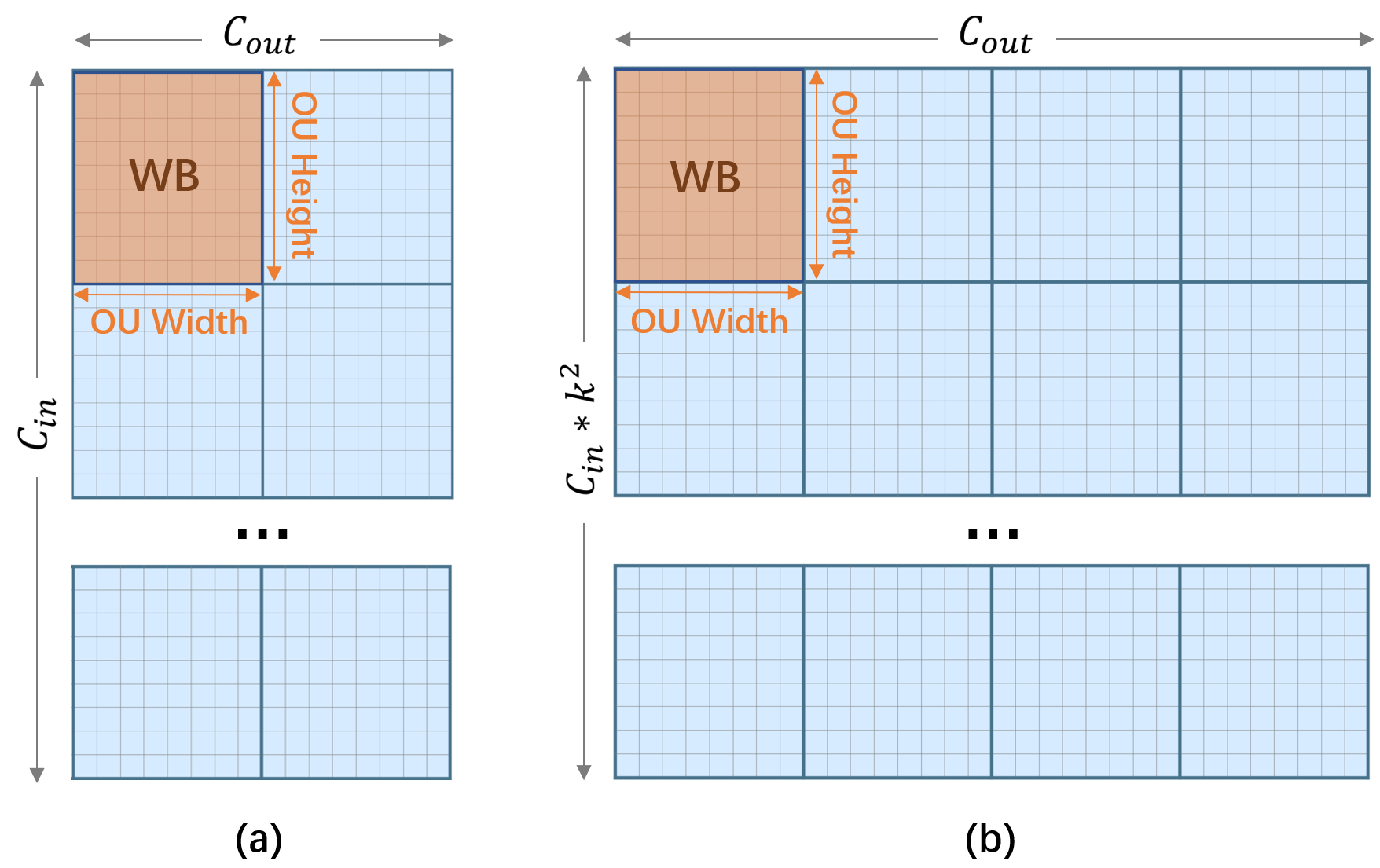}
    \vspace{-8pt}
    \caption{Partitioning weight layers into weight blocks (WBs). (a) Fully-connected layer. (b) Convolutional layer.}
    \vspace{-12pt}
    \label{weight_block_formation}
\end{figure}

To implement mixed-precision quantization at the block level, we first divide the weights of each layer into multiple 2D WBs with the same size as an OU. 
For fully-connected layers, we can directly partition the 2D weight tensor with the shape of $(C_{out}, C_{in})$, as illustrated in Fig. \ref{weight_block_formation} (a). 
However, for convolutional layers, the weights form a 4D tensor and must be flattened into 2D.  
We apply the reshaping method discussed in CSP \cite{hanson2022cascading} to reshape the convolutional layers. 
This method transforms a 4D tensor with a shape of $(C_{out}, C_{in}, k, k)$ into a 2D tensor with a shape of $((C_{in} \times k \times k), C_{out})$. Next, we divide the weights in the flattened tensor into WBs (see Fig. \ref{weight_block_formation} (b)).

Inspired by BSQ, we train the weights in their bit-level representations to exploit the bit-level sparsity. 
The bit-level representation of the weight applied in BWQ-A is defined as:

\begin{align}
     W &= sign(W) \odot \frac{s}{2^n - 1} \sum ^{n-1} _{b=0} W^{(b)}_s2^bm^{(b)},
      \label{bit-representation}
\end{align}

where $n$ denotes the number of bits, $s$ is the scaling factor, and $W^{(b)}_s$ is the $b^{th}$ bit in the binary representation, which is non-negative. 
A binary mask $m^{(b)}$ is also introduced to indicate whether a certain bit $b$ in the WB should be removed ($m^{(b)} = 0$) or retained ($m^{(b)} = 1$). 
To induce higher sparsity in the model, a WB-level group Lasso is integrated during the training process. 
The WB-level group Lasso for the $r^{th}$ layer is formulated as:
\begin{equation}
    B_{GL}(W^r) = \sum_{g=0}^{G_r-1} \sum ^{n-1} _{b=0} ||W_s ^{(g,b)}m^{(b)}||_2,
    \label{bit-GL}
\end{equation}
where $G_r$ is the total number of WBs in the $r^{th}$ layer, and $W_s^{(g, b)}$ denotes the binary representation of the weights of the $b^{th}$ bit within the $g^{th}$ WB.
Applying the WB-level group lasso is able to make a certain bit of all weight elements in the same WB zero simultaneously. 
This method regularizes the weights in each WB individually, rather than penalizing the weights in the same layer as a whole.
The overall training objective is formulated as:

\begin{equation}
    L = L_{CE} + \alpha \sum_{r=1} ^ R \frac{\#Param(W^r) \times \# Bit(W^r)}{\#Param(W^{(1:R)})}B_{GL}(W^r),
    \label{loss}
\end{equation}
where $L_{CE}$ is the cross-entropy loss, $R$ is the number of layers of the model, and $\alpha$ is the regularization strength. 
To minimize the total number of bits in the model, the loss function includes coefficients for each WB-level group Lasso.
These coefficients impose greater penalties on layers with a higher number of bits.

\begin{figure}
    \centering
    \includegraphics[width=0.48\textwidth]{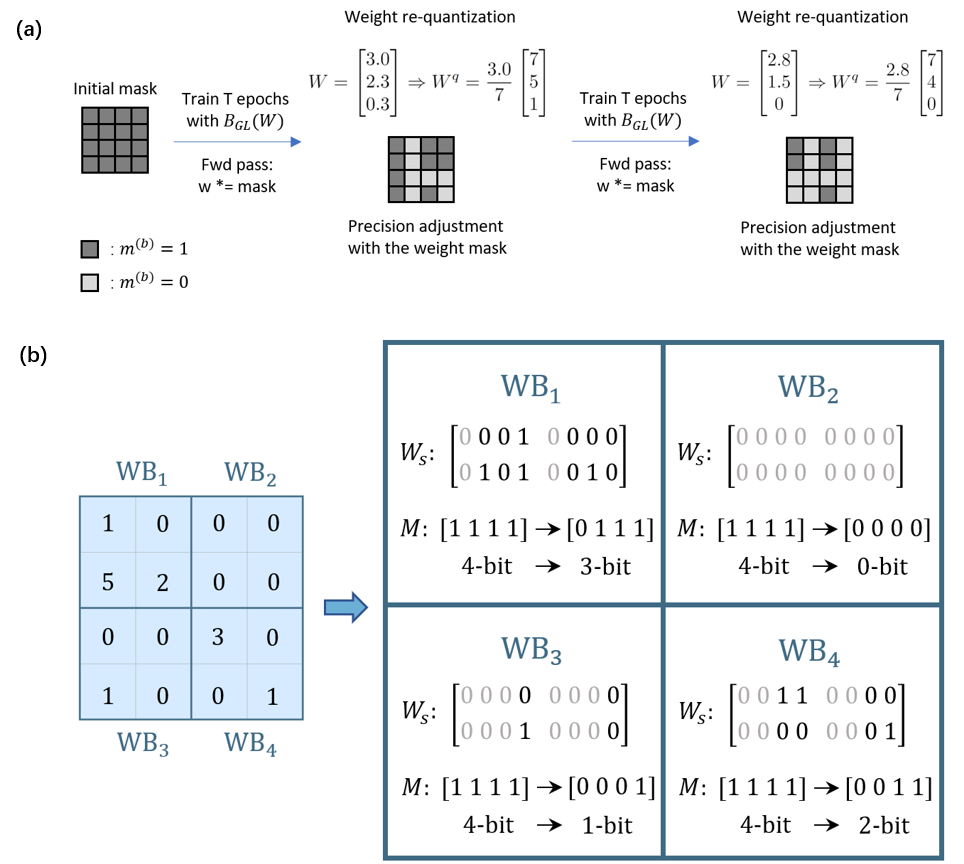}
    \caption{(a) The quantization-aware-training scheme of BWQ-A. (b) Block-wise precision adjustment operation.}
    \label{re-quant}
\end{figure}

To enhance the model's resilience against quantization noises, we conduct regular re-quantization followed by adjusting the precision in a block-wise manner at specific intervals throughout the training process. 
The overall weight compression scheme of BWQ-A is illustrated in Fig. \ref{re-quant} (a).
In the re-quantization process, we convert each bit of the weights into exact binary values.
The precision adjustment scheme in BWQ-A, which is depicted in Fig. \ref{re-quant} (b), conducts the removal of the zero-valued bits in a block-wise manner. 
The initial precision for all WBs is set to 8-bit (in the simplified example of Fig. \ref{re-quant} (b), the WBs are initially set to 4-bit), and the initial value of the binary bit mask $m^{(b)}$ for every bit of the weights is 1.
During the precision adjustment process, we check each bit of $W_s$ within a WB from the most significant bit (MSB) down to the least significant bit (LSB). 
If a certain bit for all the weight elements within a WB only contains zeros, then the bit could be removed in this block by setting the corresponding bit mask to 0. 
The bit value checking stops until we encounter the first non-zero bit. 
In this way, the WBs are assigned different precisions according to their individual contribution towards the training objective.
As illustrated in Fig. \ref{re-quant} (a), in the forward pass, the weights are updated with the product of the weight tensor and the mask tensor. 
Therefore, the pruned bits of the weights will remain zero in the subsequent training epochs.
This ensures that the sparsity of the model is non-decreasing throughout the training process.

Our compression objective is to achieve the highest possible weight compression ratio while ensuring that the quantized model's accuracy does not drop by more than $1\%$. Therefore, we gradually increase the regularization strength by a fixed interval $\Delta_{\alpha}$ until the accuracy loss is larger than $1\%$.


\begin{algorithm} 
        \caption{Quantization scheme of BWQ-A.}
	\begin{algorithmic}[1]
		\REQUIRE $\Delta_{\alpha}$, Init\_Act\_Precision, Init\_Weight\_Precision
        \STATE Act\_Precision $\gets$ Init\_Act\_Precision
        \STATE $\alpha \gets 0$
        \STATE M $\gets$ 1  /*Initialize the binary mask*/
        \STATE $W_{fp} \gets$ Random\_Weight\_Init()
        \STATE $W_b \gets$ Bit\_Representation($W_{fp}$, Init\_Weight\_Precision)
        \STATE /*Determine the weight regularization strength*/
        \WHILE{acc\_loss $\leq$ 1\%}
            \STATE $\alpha \gets \alpha + \Delta_{\alpha}$
            \FOR{epoch=1,..., Total\_Training\_Epoch}
                \STATE train($W_b$, M, $\alpha$, Act\_Precision)
                \IF{epoch in Quant\_Epochs}
                    \STATE $W_b \gets$ Quant($W_b$)
                    \STATE M $\gets$ Prec\_Adjust($W_b$)
                \ENDIF
            \ENDFOR
        \ENDWHILE
        \STATE /*Determine the activation precision*/
        \WHILE{acc\_loss $\leq$ 1\%}
            \STATE Act\_Precision $\gets$  Act\_Precision - 1
            \FOR{epoch=1,..., Total\_Training\_Epoch}
                \STATE train($W_b$, M, $\alpha$, Act\_Precision)
                \IF{epoch in Quant\_Epochs}
                    \STATE $W_b \gets$ Quant($W_b$)
                    \STATE M $\gets$ Prec\_Adjust($W_b$)
                \ENDIF
            \ENDFOR
        \ENDWHILE
        \RETURN $W_b$, M, Act\_Precision
	\end{algorithmic} 
    \label{BWQ-A-algo}
\end{algorithm}

\subsection{Activation Compression}
\begin{equation}
    y = PACT(x) = 0.5(|x| - |x - \beta| + \beta) = \left\{
             \begin{array}{lr}
             0, & x \in (-\infty, 0) \\
             x, & x \in [0, \beta)\\
             \beta, & x \in [\beta, +\infty)
             \end{array}
\right.
    \label{PACT}
\end{equation}

To minimize the end-to-end latency of the accelerator, BWQ-A performs activation quantization as well, as reducing the activation bit-width effectively decreases the buffer's read latency.
Weight compression and activation compression are two orthogonal processes in BWQ-A, and we predetermine the activation bit-width before each training process. 
To achieve low-bit-width activations, we apply the Parameterized Clipping Activation Function (PACT) proposed in \cite{choi2018pact} prior to activation quantization. 
As shown in Equation \ref{PACT}, 
PACT removes outliers of the activation values and constrains the activations to a relatively small range, thereby decreasing the quantization noise. 
We first determine the weight quantization scheme with \SI{8}{\bit} activations and then gradually decrease the activation precision until the accuracy degradation exceeds $1\%$.
The overall quantization scheme of BWQ-A is illustrated in Algorithm \ref{BWQ-A-algo}.


\section{BWQ-H: Hardware Acceleration Of BWQ-A With ReRAM-PIM} \label{BWQ-H}
This section describes the implementation of BWQ-H, which is designed for the efficient inference of BWQ-A models. 
BWQ-H is a practical ReRAM-based PIM accelerator that adopts the OU-based operational scheme. 
While the OU-based scheme limits the parallelism of the ReRAM crossbars, the confined number of simultaneously activated WLs in an OU allows for a significant reduction in the required ADC precision.
Since ADC power consumption increases exponentially with its precision, accelerators with the OU-based operation scheme can achieve more energy-efficient performance by utilizing low-precision ADCs.

\begin{figure}
    \centering
    \includegraphics[width=0.48\textwidth]{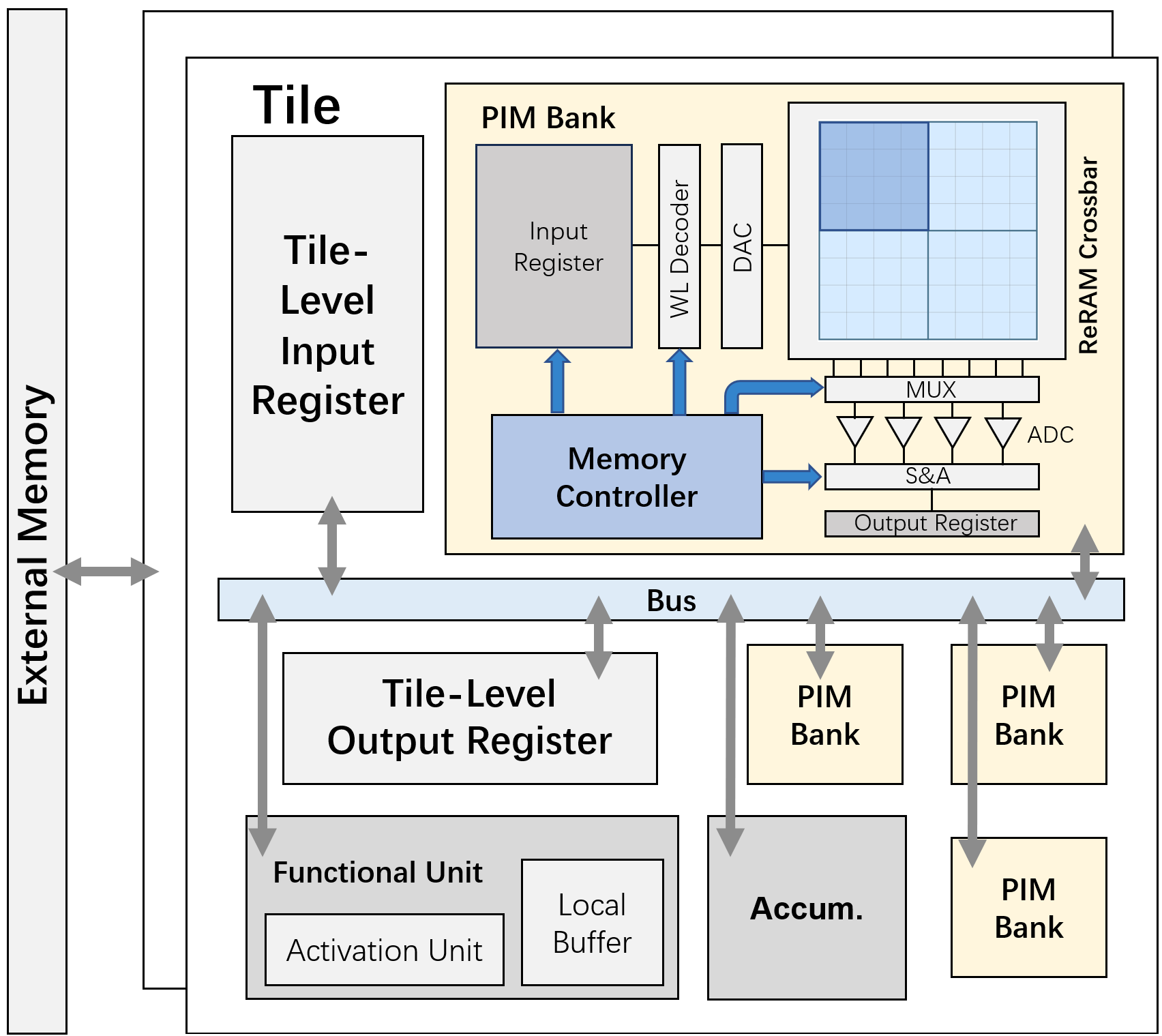}
    \caption{Architecture of BWQ-H. 
    The grey arrows represent the dataflow, while the blue arrows denote the control flow of the memory controller.}
    \label{BWQ-H-arch}
\end{figure}

The overall architecture of BWQ-H is illustrated in Fig. \ref{BWQ-H-arch}.  
The accelerator comprises multiple tiles connected through a Network-on-Chip (NoC) and external memory. 
Within each tile, there are several PIM banks, accumulation units, tile-level input/output registers, functional units, and the local bus. 
Each PIM bank includes a ReRAM crossbar, input/output register, WL decoder, DACs, MUX, ADCs, S\&A units, and the memory controller.

In the remainder of this section, we first introduce a precision-aware weight mapping method that increases the utilization of the ReRAM crossbar arrays for the mixed-precision quantized models. 
Next, we present the design of a memory controller, which generates control signals for the peripheral circuits to enable the computation with mixed-precision weights on the same crossbar.

\subsection{Precision-Aware Weight Mapping}


\begin{figure}
    \centering
    \includegraphics[width=0.48\textwidth]{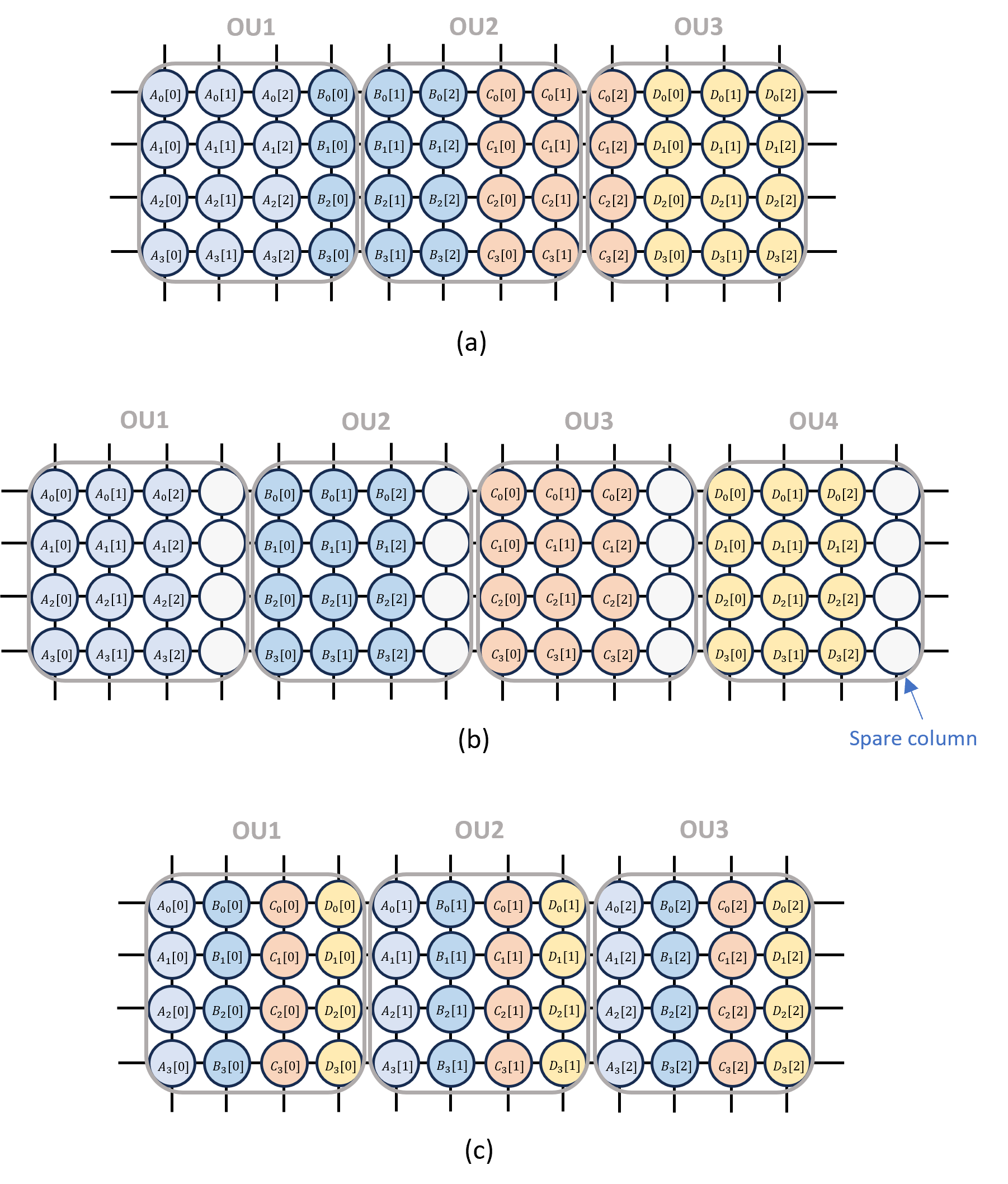}
    \caption{Example of different weight mapping schemes. Cells with the same color represent the same weight vector, $V_i[n]$ denotes the $n^{th}$ bit of the $i^{th}$ element in vector $V$. 
    (a) Conventional mapping scheme, in which bits of the same weight element may span different OUs. (b) A mapping scheme that constrains the bits of the same weight element to be mapped on the same OU. The blank cells represent spare columns. (c) Precision-aware mapping scheme.}
    \label{OU_mapping}
\end{figure}

Fig. \ref{OU_mapping} provides an illustration of different weight mapping schemes. 
In this example, a $4 \times 4$ WB is mapped to the crossbar consisting of $4 \times 4$ OUs. 
Within the WB, there are 4 3-bit weight vectors A, B, C, and D.
Fig. \ref{OU_mapping} (a) illustrates the traditional weight mapping method. 
Here, different bits from the same weight vector are mapped in consecutive columns of the crossbar array. 
For our mixed-precision quantized models, however, this mapping method will lead to an induced peripheral circuit overhead or low OU utilization if the number of columns in an OU is not divisible by the weight precision in a certain block, as shown in Fig. \ref{OU_mapping} (a) and (b). 
In Fig. \ref{OU_mapping} (a), since the second weight vector $B$ spans two OUs, the computation results with $B[0]$ on OU1 and the results with $B[1]$ and $B[2]$ on OU2 should be accumulated. 
Thus, complex indexing control logic for shift-and-add (S\&A) units is required to sum up the computation results with bits of the same weight vector from different OUs activated in different clock cycles. 
This can cause additional overhead for the peripheral circuitry and result in increased computation latency.  
To avoid the overhead associated with the complex control logic, one alternative is to only allow the bits of the same weight to be mapped within the same OU, as depicted in Figure \ref{OU_mapping} (b).
However, this can lead to low memory utilization of the OUs, thus leading to throughput reduction. In the given example, each OU would contain a spare column, resulting in a 25\% reduction in OU throughput.

To enable efficient computation of the mixed-precision quantized models, we propose a precision-aware weight mapping scheme, which is illustrated in Fig. \ref{OU_mapping} (c). Instead of mapping different bits of the same weight vector in consecutive columns, we assign them to different OUs. 
Within an OU, we map the same bit of different weights in the $4 \times 4$ WB. 
The computation results with different bits of the same weight vector are accumulated by the S\&A units from different OUs.
This precision-aware mapping scheme eliminates the need for additional peripheral circuits and is able to achieve $100\%$ memory utilization within an OU.

\begin{figure}[h]
    \centering
    \includegraphics[width=0.45\textwidth]{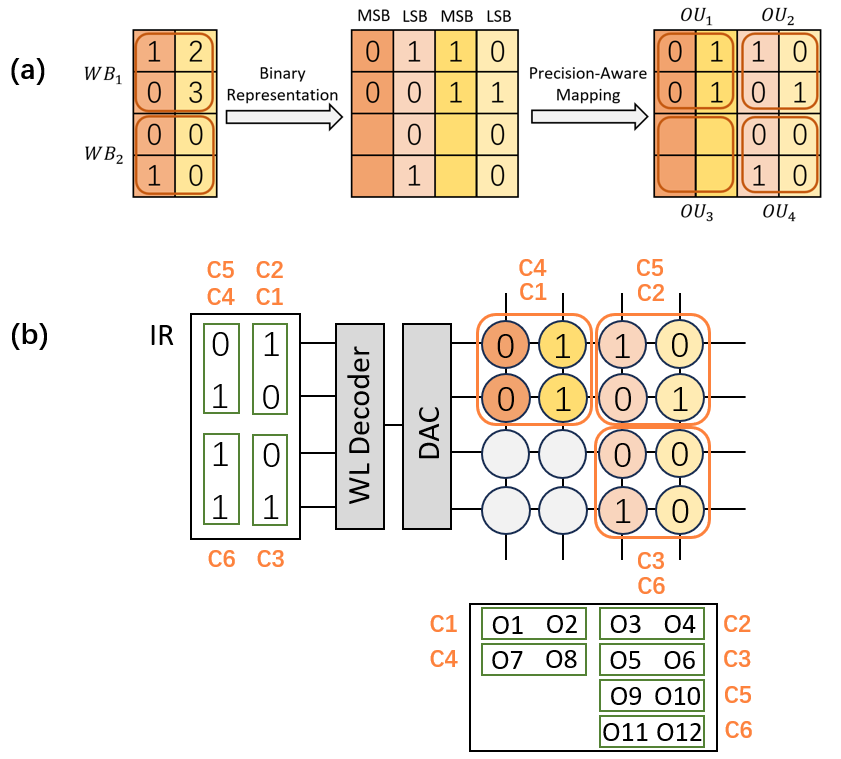}
    \caption{(a) Mapping two WBs with different precisions to a 4 $\times$ 4 crossbar array with the precision-aware mapping scheme. (b) An example of mixed-precision computation on BWQ-H. $C_i$ denote the $i^{th}$ cycle.}
    \label{mp-computation}
\end{figure}

Fig. \ref{mp-computation} (a) shows an example of mapping multiple WBs with different precisions onto the crossbar with the aforementioned scheme. 
This is a simple example of mapping a weight tensor that only contains two $2 \times 2$ WBs. 
Note that if the total bit-width of each row of WBs varies, there may exist spare OUs in the crossbar (as illustrated by $OU_3$ in Fig. \ref{mp-computation} (a).
However, this will not impact the throughput of the OUs or result in increased power consumption.
This is due to the fact that within one clock cycle, only one OU can be activated, and the computation with unused OUs will be skipped. 
Moreover, the capacity of the peripherals in each PIM bank is designed to handle the operations of a single OU. 
Owing to the high density of the ReRAM crossbars in contrast to the peripheral circuits, the spare OUs would only result in negligible area overhead \cite{shafiee2016isaac}.


\subsection{Memory Controller Design}
To enable the computation with mixed-precision weights on the same crossbar, a memory controller is required to generate control signals for the peripheral circuits.
The memory controller contains a look-up-table (LUT) to store the bit-width of all the WBs in one PIM bank, and generates control signals to the WL decoder, multiplexer (MUX), the S\&A units, and the input register (IR) at the PIM bank level. 

The control logic of the memory controller is illustrated in Algorithm \ref{ctrl-alg}.
To perform OU-based computation with the mixed-precision WBs, the controller 
(1) generates the row and column addresses for the current OU, and sends the addresses to the WL decoder and the MUX so that the corresponding rows and columns are activated;
(2) produces a skip signal for the S\&A upon completing the computation across multiple OUs related to one WB, so that the computation results of different WBs are not accumulated;
(3) generates an enable signal for the IR once the computation for a row of WBs is finished, so that the corresponding activation data for the next WB row can be fetched from the register and sent to the activated WLs.
The control signal flow is denoted by the blue arrows in Fig. \ref{BWQ-H-arch}.

Fig. \ref{mp-computation} (b) shows an example of how BWQ-H conducts mixed-precision computations coordinated by the memory controller. 
In the first cycle C1, the activation data stored in the upper-right section of the IR is sent to the 1st and 2nd WLs, and performs the multiply-accumulate (MAC) operation with $OU_1$. 
This results in the outputs of O1 and O2. 
To reuse the activation data, in C2, we perform the MAC operation with $OU_2$, and generate O3 and O4. 
Since OU1 and OU2 store different bits of the same WB, the outputs in C1 and C2 are accumulated by the S\&A units.
Next, in C3, we fetch the activation data in the bottom-right section of the IR and send it to the 3rd and 4th WLs. 
As $OU_3$ is a spare OU, we skip its computation and proceed to $OU_4$.
In C3, the S\&A units receive a skip signal from the memory controller, preventing O5 and O6 from being accumulated with previous outputs. 
To complete the computation presented in this example, a total of 6 cycles are required. The subsequent computation sequence is depicted in Fig. \ref{mp-computation} (b).

\begin{algorithm} 
        \caption{Control logic of the memory controller.}
	\begin{algorithmic}[1]
        \REQUIRE $C_{out}, C_{in}$, k, OU\_Width, \\
            OU\_Height, Bitwidth\_Table
        \STATE Num\_Hblock $\gets$ Ceil($C_{out}$ / OU\_Width) 
        \STATE Num\_Vblock $\gets$ Ceil($C_{in}$ * k * k / OU\_Height)
        \STATE Col\_Start\_Idx $\gets 0$
        \STATE Fetch\_Next $\gets 0$
        \FOR{i = 0, 1, ..., Num\_Hblock - 1}
            \STATE Fetch\_Next $\gets 0$
            \FOR{j = 0, 1, ..., Num\_Vblock - 1}
                \STATE Weight\_Precision $\gets$ BW\_Table[i-1][j-1]
                \IF{Weight\_Precision $\neq 0$}
                    \STATE Psum $\gets 0$
                    \STATE Activated\_Rows $\gets$ [(i-1) * OU\_Height : \\i * OU\_Height]
                    \FOR{k = 0, 1, ..., Weight\_Precision - 1}
                        \STATE Activated\_Cols $\gets$ [Col\_Start\_Idx : \\    
                            Col\_Start\_Idx + OU\_Width]
                        \STATE Psum $\gets$ Shift\_Left(Psum) + \\
                            Current\_ADC\_Output
                        \STATE Col\_Start\_Idx $\gets$ Col\_Start\_Idx + OU\_Width
                    \ENDFOR
                \ENDIF
                \IF{j == Num\_Vblock - 1}
                    \STATE Fetch\_Next $\gets 1$
                \ENDIF
            \ENDFOR
        \ENDFOR
	\end{algorithmic} 
    \label{ctrl-alg}
\end{algorithm}

\section{Evaluation Methodology} \label{exp-setup}
To assess BWQ's efficacy, we examine the impact of its two components, BWQ-A and BWQ-H, both separately and in combination. 
First, we verify the influence of BWQ-A on the model's performance and compression ratio. 
Then, we evaluate BWQ-H's inference latency and energy overhead with the quantized models obtained by BWQ-A to examine the effectiveness of the proposed co-design scheme.
\subsection{Algorithm Performance Validation}
BWQ-A is compared against the floating-point baseline models and BSQ models in terms of accuracy and model compression ratio. 
We consider a spectrum of representative models on CIFAR-10, CIFAR-100 \cite{krizhevsky2009learning} and ImageNet \cite{imagenet15russakovsky} datasets. 
For the CIFAR experiments, ResNet-20, ResNet-18, ResNet-34 \cite{he2016deep}, VGG16-BN, VGG19-BN \cite{simonyan2014very}, MobileNetV2 \cite{sandler2018mobilenetv2} and DenseNet-121 \cite{huang2017densely} are examined. 
For the evaluations on ImageNet, we use ResNet-34 and DenseNet-121.

In the CIFAR experiments, the floating-point baseline models are trained for 200 epochs with SGD optimizer with 0.9 momentum and 1e-4 weight decay.
The initial learning rate is 0.1, and a cosine annealing learning rate scheduler is utilized. 
The training for BWQ-A uses the same hyperparameters as the baseline models, except that BWQ-A is trained for 650 epochs because it incorporates a quantization-aware-training approach. 
This extension of training epochs proves advantageous as it facilitates the models' adaptation to the quantization noises.
The number of training epochs is comparable with the total training epochs (training and fine-tuning) used in previous studies such as BSQ \cite{yang2021bsq} and CSQ \cite{xiao2022csq}.
The VGG models are re-quantized every 200 epochs, while the ResNet, DenseNet, and MobileNet models are re-quantized every 300 epochs. 
A final re-quantization and precision adjustment operation is conducted after the training process, finalizing the precisions of the WBs.

For the ImageNet experiments, we use the pretrained models from torchvision \cite{torchvision2016} as the floating-point baseline models.
The BWQ-A models are trained for 90 epochs with SGD optimizer with 0.9 momentum.
We set the initial learning rate as 0.001, and use a cosine annealing learning rate scheduler.
A weight decay parameter of 1e-4 is incorporated.
The models are re-quantized at the 60th and the 90th epoch.
All the training processes are conducted using two A5000 GPUs with distributed data parallel.

\subsection{Architecture Modelling and Comparison}
For hardware performance, we compare BWQ-H with several representative ReRAM-based accelerators, including a baseline architecture ISAAC \cite{shafiee2016isaac} with no model compression, SRE \cite{yang2019sparse}, BSQ \cite{yang2021bsq} and SME\cite{liu2021sme}, which perform model compression with HW-based, SW-based and HW/SW co-design approaches, respectively. 
The detailed operational principles of SRE and SME are discussed in Section \ref{BWQ-H results}. 
BSQ's hardware performance on a ReRAM-based accelerator is simulated by modifying BWQ-H and following the same simulation method as BWQ-H.
We modified the MNSIM \cite{zhu2020mnsim} simulator to evaluate the performance of BWQ-H. 
For the analysis of the overhead for implementing the BWQ-A model with the OU-based operation scheme, we build an RTL model in Verilog for the memory controller and synthesize the model using Synopses Design Compiler with TSMC commercial 28nm standard library.
Table \ref{HW config} shows the hardware configuration to evaluate the BWQ-H architecture.

\begin{table}
    \centering
    \caption{Hardware Configuration.}
    \begin{tabular}{|c|c|c|}
        \hline
        \multicolumn{3}{|c|}{\textbf{BWQ-H configuration} (1.2 GHz)} \\ \hline
        \textbf{Component} & \textbf{Spec} & \textbf{Power} \\ \hline
        Memristor & size: 128$\times$128;  & \multirow{2}{*}{0.89W} \\ 
        array & bit-per-cell: 1; OU size: 8$\times$9 &  \\ \hline
        DAC & resolution: 1-bit & 0.36W \\ \hline
        ADC & resolution: 4-bit & 23.22W \\ \hline
        Buffer & bitwidth: 64-bit & 0.59W \\ \hline
        Memory & \multirow{2}{*}{tech node: 28nm} & \multirow{2}{*}{92.8mW} \\ 
        controller & & \\ \hline
        Other digital & S+A: 4 per PIM Bank; & \multirow{2}{*}{92.6mW} \\ 
        components& IR: 2KB; OR: 256B & \\ \hline
        \textbf{Chip total} & & 25.25W \\ \hline
    \end{tabular}
    \label{HW config}
\end{table}


\section{Experimental Results} \label{results}
\begin{figure*}
    \centering
    \includegraphics[width=0.98\textwidth]{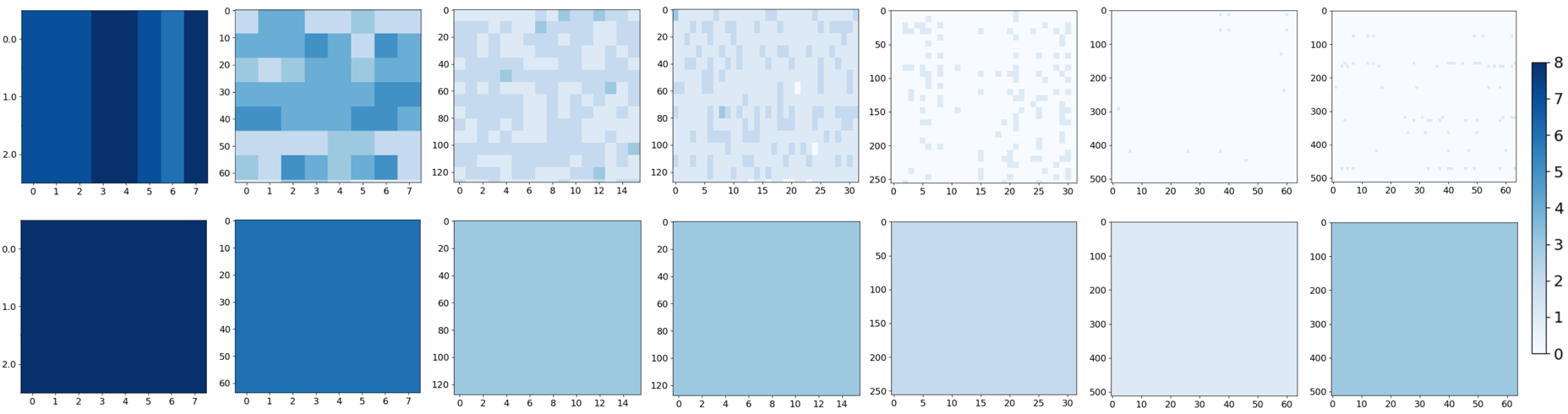}
    \caption{Quantization results of 7 representative layers of VGG19-BN trained on CIFAR-10. The first and the second lines present the quantization results of BWQ-A and BSQ, respectively. 
    Each column contains two bit-width maps corresponding to the same weight layer.
    The X-axis and Y-axis represent the index of the WBs in the $C_{out}$ dimension and the $C_{in}$ (or $C_{in} \times k \times k$) dimension, respectively. 
    A colorbar is presented on the right. 
    The specific layers are: ``features.0", ``features.4", ``features.13", ``features.18", ``features.26", ``features.43", ``features.64".}
    \label{quant-results}
\end{figure*}

\begin{table*}[t]
  \centering
  \caption{Performance Comparison of the Compressed Models.}
  \label{combinedtable}
  \begin{tabular}{l l r r r r r r r r}
    \toprule
    \multirow{2}{*}{Dataset} 
    & \multirow{2}{*}{Model} 
    & \multirow{2}{*}{\#Param (M)} 
    & \multirow{2}{*}{Baseline Acc (\%)} 
    & \multicolumn{3}{c}{BSQ}
    & \multicolumn{3}{c}{BWQ-A}\\
    \cmidrule(r){5-7} \cmidrule(r){8-10}
    &  &  &  & Act. Prec. & Acc (\%) & Comp. ($\times$) & Act. Prec. & Acc (\%) & Comp. ($\times$) \\
    \midrule
    \multirow{6}{*}{CIFAR-10} & ResNet-18 & 11.17 & 95.38 & 4 & 95.38 & 26.05 & 3 & 94.59 & 56.46 \\
     & ResNet-34 & 21.28 & 95.61 & 4 & 94.66 & 83.86 & 4 & 94.73 & 117.52 \\
     & VGG16-BN & 33.65 & 92.60 & 3 & 92.45 & 26.59 & 3 & 92.60 & 136.01 \\
     & VGG19-BN & 38.96 & 92.94 & 3 & 92.84 & 28.15 & 3 & 91.96 & 443.01 \\
     & ResNet-20 & 0.27 & 92.61 & 3 & 92.16 & 13.76 & 3 & 92.07 & 16.04 \\
     & MobileNetV2 & 2.30 & 94.43 & 4 & 94.24 & 5.73 & 3 & 93.54 & 47.34 \\
     \hline
    \addlinespace
    \multirow{6}{*}{CIFAR-100} & ResNet18 & 11.22 & 75.61 & 4 & 76.35 & 6.21 & 4 & 75.31 & 45.97 \\
     & ResNet34 & 21.33 & 76.41 & 4 & 76.37 & 33.40 & 4 & 76.16 & 63.93 \\
     & VGG16-BN & 34.02 & 72.93 & 4 & 72.37 & 19.88 & 3 & 72.87 & 50.42 \\
     & VGG19-BN & 39.33 & 72.94 & 4 & 72.05 & 23.09 & 3 & 72.04 & 78.56 \\
     & MobileNetV2 & 2.41 & 75.54 & 8 & 75.13 & 6.22 & 4 & 74.57 & 18.35 \\
     & DenseNet121 & 7.00 & 75.99 & 5 & 75.18 & 7.26 & 4 & 75.07 & 23.65 \\
     \hline
    \addlinespace
    \multirow{2}{*}{ImageNet} & ResNet34 & 21.80 & 73.55 & 4 & 72.62 & 9.48 & 4 & 72.56 & 13.55 \\
     & DenseNet121 & 7.98 & 74.65 & 5 & 73.70 & 7.47 & 4 & 73.68 & 12.56 \\
    \bottomrule
  \end{tabular}
  \label{BWQ-A-results}
\end{table*}

\subsection{Algorithm}
\begin{figure}
    \centering
    \includegraphics[width=0.46\textwidth]{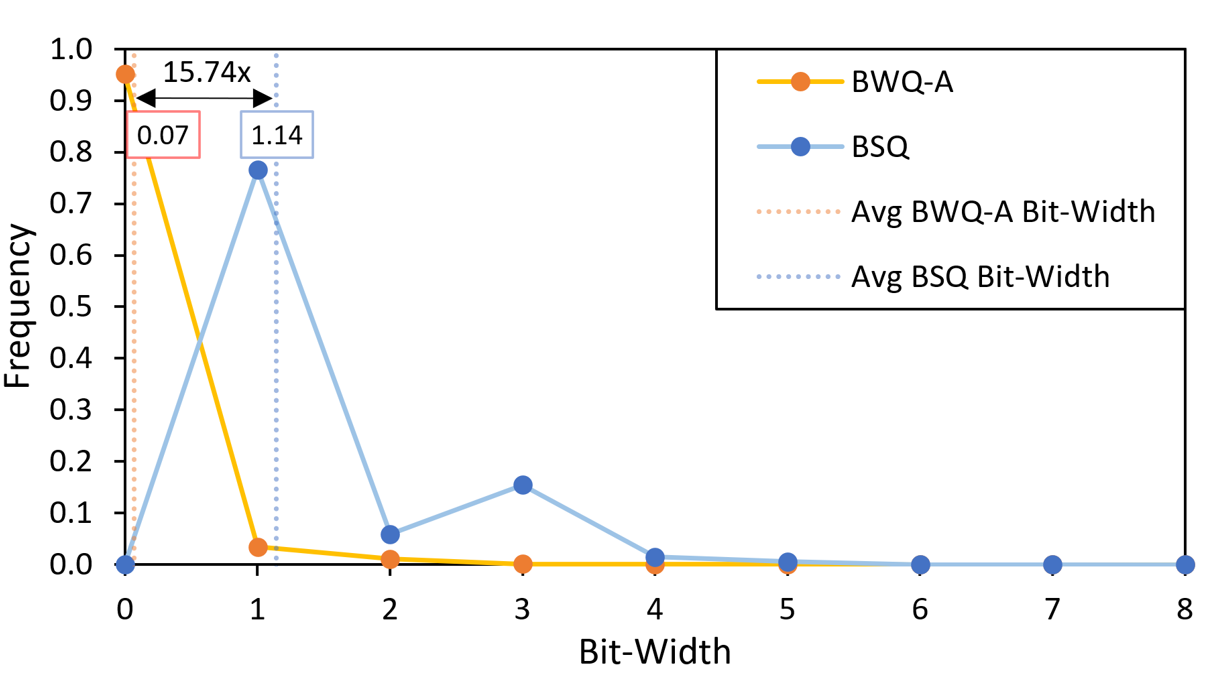}
    \caption{Bit-width distribution of the VGG19-BN model trained on CIFAR-10 with BWQ-A and BSQ.}
    \label{bw-dist}
\end{figure}
In this section, we compare the performance of BWQ-A with 32-bit floating-point baseline models and BSQ. 
The experimental results on CIFAR-10, CIFAR-100, and ImageNet are presented in Table \ref{BWQ-A-results}. 
Here, `Act. Prec.' refers to activation precision, `Acc' is the test accuracy, and `Comp' is the model compression ratio. 
Typically, BWQ-A is able to achieve a higher weight and activation compression ratio with more redundant models. 
On the CIFAR-10 dataset, BWQ-A achieves an average (geometric mean) weight compression ratio of $81.98\times$ and an average activation compression ratio of $10.17\times$, all while maintaining an accuracy degradation within 1\% compared to the floating-point baseline models. 
Image classification tasks on CIFAR-100 and ImageNet are more complicated.
However, on average, BWQ-A is still able to attain a $41.42 \times / 13.05 \times$ weight compression ratio and an $8.81 \times / 8.00 \times$ activation compression ratio, respectively. 

Compared with BSQ, BWQ-A is able to achieve a higher model compression ratio with similar accuracy, as BWQ-A leverages a finer mixed-precision quantization granularity.
Fig. \ref{quant-results} shows the quantization results of 7 layers of VGG19-BN trained on CIFAR-10 with BSQ and BWQ-A. 
The first and the second rows of heatmaps present the quantization results of BWQ-A and BSQ, respectively. 
Each column contains two bit-width maps corresponding to the same weight layer.
According to Fig. \ref{quant-results}, in BWQ-A, as the WBs contribute differently to the training objective, they can be assigned flexible precisions based on their individual significance. 
In contrast, BSQ restricts the weights in the same layer to have the same precision. 
In BSQ, although many of the WBs may have low significance, all of the weights are designated with the maximum bit-width of the WB for that specific layer. 
The effects of this issue become more pronounced in deeper layers that contain a larger number of parameters.
This drawback prevents BSQ from achieving the optimal trade-off between compression ratio and accuracy, rendering BSQ more resource-intensive when deployed on ReRAM-based accelerators. 

Fig. \ref{quant-results} (c) presents the bit-width distribution of the entire VGG19-BN model trained on CIFAR-10. We observe an evident shift in the distribution of these two model compression schemes. 
Notably, BWQ-A achieves a substantial reduction in average bit-width compared to BSQ, with an average $15.74 \times$ lower bit-width. 

\begin{figure*}
    \centering
        \includegraphics[width=1\textwidth]{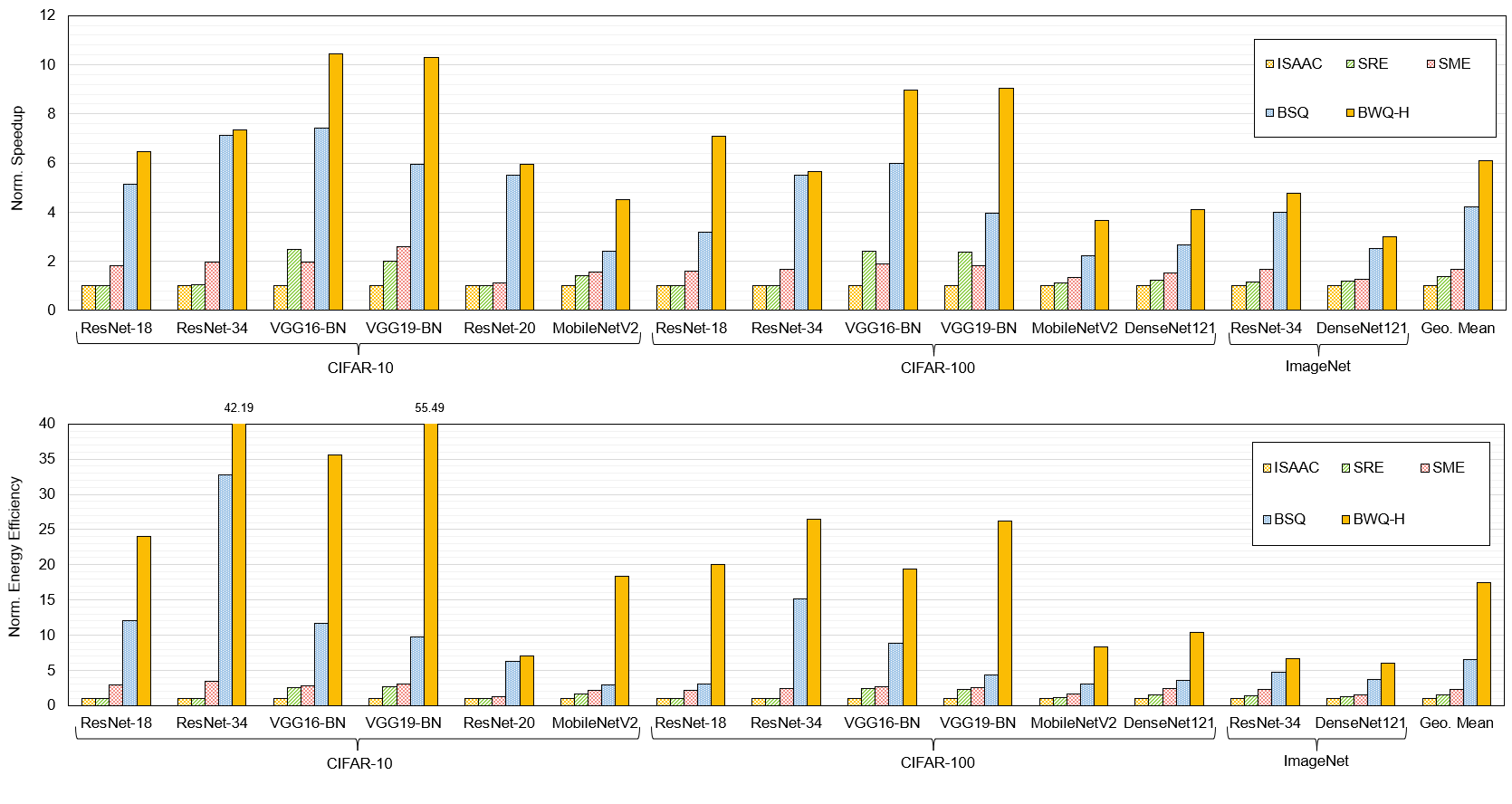}
    \caption{The normalized speedup and energy efficiency (over ISAAC) of BWQ-H and other baseline accelerators.}
    \label{latency-energy}
\end{figure*}

\subsection{Accelerator} \label{BWQ-H results}
\begin{figure}
    \centering
    \includegraphics[width=0.48\textwidth]{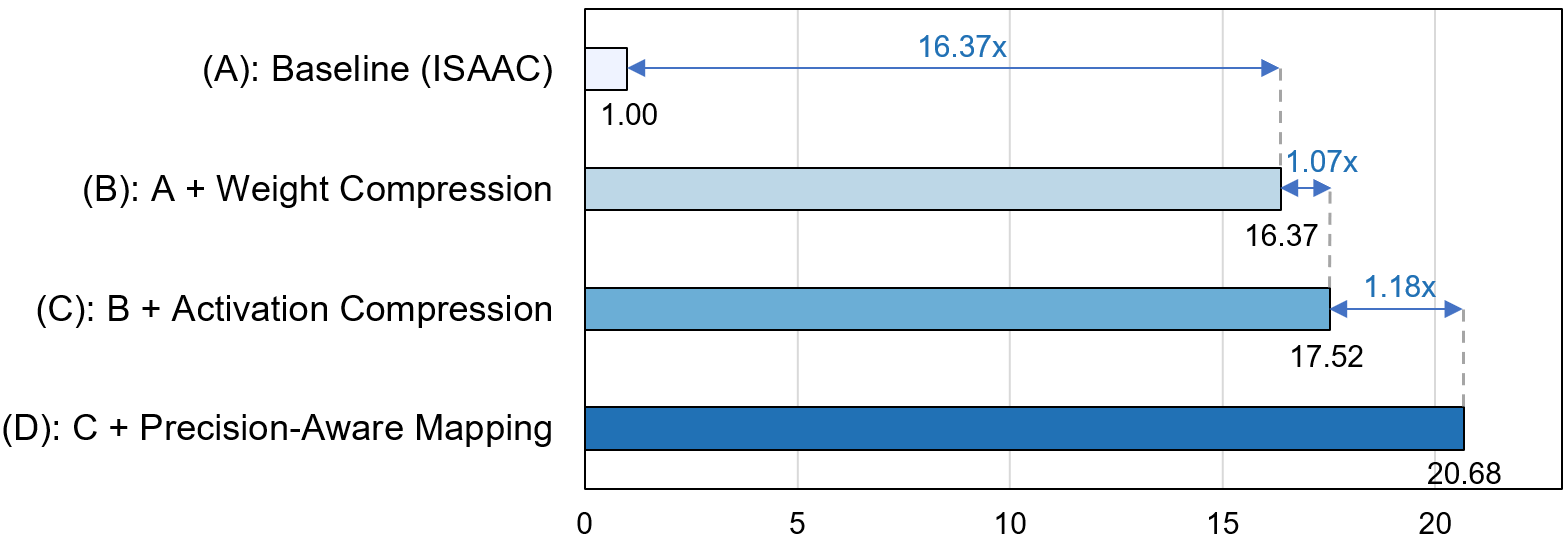}
    \caption{Breakdown analysis for BWQ-H's energy saving over ISAAC.}
    \vspace{-6pt}
    \label{energy-breakdown}
\end{figure}

Fig. \ref{latency-energy} shows the normalized speedup and energy efficiency of all accelerators considered in this work normalized with respect to ISAAC. 
The performance of the accelerators is simulated under the OU-based operation scheme to ensure high computational accuracy and manageable peripheral overhead.
Our proposed architecture BWQ-H is able to achieve the highest speedup and energy efficiency for all models and datasets considered in this work. 
On average, BWQ-H achieves $6.08 \times$ speedup and $17.47 \times$ energy saving over ISAAC. 
Fig. \ref{energy-breakdown} illustrates the breakdown of BWQ-H's energy saving over ISAAC. 
The primary source of energy saving is attributed to the high weight compression ratio achieved through BWQ-A, while activation compression and the precision-aware mapping scheme also contribute to reducing BWQ-H's energy consumption. 
In the OU-based operation scheme, the increased number of cycles needed to complete the computation with the entire crossbar leads to a significant rise in ADC energy, which becomes the dominant part of energy consumption in the ReRAM-based accelerators. 
Therefore, applying models with a high weight compression ratio can effectively reduce the number of computation cycles, thus greatly enhancing energy efficiency.

A notable observation from Fig. \ref{latency-energy} is that on both CIFAR-10 and CIFAR-100 datasets, despite achieving a significantly higher weight compression ratio with BWQ-A, BWQ-H's speedup for VGG19-BN is not higher than that of VGG16-BN. 
This is due to the fact that BWQ-H optimizes specific parts of the system, and there exists a speedup limit determined by the latency of the unoptimized components. 
BWQ-H is reaching the speedup limit on both models. 
However, being a deeper model, VGG19-BN introduces more layers with large input feature maps. As a result, the latency of relatively lower optimized parts, such as buffers and accumulation circuits, is increased. 

Since the baselines represent different types of architectures, we discuss the comparison between BWQ-H's design with each baseline architecture separately.

\textbf{ISAAC}: 
To guarantee high computational accuracy, the 2-bit ReRAM cells assumed in ISAAC are replaced by 1-bit cells in our simulations.
In ISAAC, both weights and activations are represented with 16-bit. 
While this configuration delivers nearly lossless accuracy, it's deemed excessive and resource-demanding for edge devices.
The primary reason for BWQ-H's superior performance over ISAAC is its superior weight and activation compression rates. 
On average, BWQ-H achieves a speedup of $6.08\times$ and saves $17.47\times$ more energy than ISAAC.

\textbf{SRE}:
SRE applies a HW-based optimization approach, exploring very fine-grained structured sparsity on the OU-row level from both the weight and the activation side. 
This method omits zero-valued OU rows, replacing them with subsequent non-zero values.
However, this optimization approach only leverages the inherent sparsity of neural networks but does not proactively compress the models.
Thus, its weight compression performance is much lower than what BWQ-A can achieve. 
For example, the highest average compression ratio that SRE can achieve is about $10 \times$, and this is accomplished with the smallest $2 \times 2$ OUs.
For $9 \times 8$ OUs, the compression ratio is only about $3.3 \times$. 
On the other hand, to leverage fine-grained sparsity, SRE imposes substantial peripheral overheads for indexing control, including index storage and complex indexing logic.
Therefore, BWQ-H is able to achieve $4.44\times$ average speedup and $11.98\times$ energy saving over SRE. 

\textbf{SME}:
SME proposes a HW/SW co-design approach to design a DNN acceleration framework for ReRAM devices. 
At the algorithm level, SME performs post-training-quantization to compress the model. 
In the proposed quantization scheme, at most 3 consecutive bits of the weights can be non-zero. 
To leverage the bit-level sparsity obtained from the compression algorithm, SME proposes a bit-wise inter-crossbar mapping scheme and a squeeze-out scheme.
In the mapping scheme, the different bits of a weight matrix are mapped to different crossbars. 
In the squeeze-out process, if a crossbar row representing the LSB only contains zeros, then the zeros are squeezed out and the row is replaced with the non-zero values from the other crossbars with a right-shift.
This requires doubling the corresponding activation values. 
However, SME's quantization scheme can significantly compromise model accuracy due to its constraints on the number of non-zero bits. 
Moreover, the mapping and squeeze-out methods only leverage the sparsity to a very limited extent. 
Only if an entire crossbar row consists of zero-weight values then the data in that row can be squeezed out. 
Thus, the de facto model compression ratio is low when large crossbars are utilized.
Since SME requires doubling the activation data on specific rows, it also induces additional overhead to record the index of the activation data that is required to be doubled. 
Therefore, BWQ-H is able to achieve $3.66\times$ average speedup and $7.65\times$ energy saving over SME. 

\textbf{BSQ}:
As introduced in Section \ref{mpq}, BSQ performs layer-wise mixed-precision quantization to compress both the weights and activations. 
Since all the weights within a layer have a uniform precision, BSQ doesn't necessitate a memory controller to track the bit-width for each WB as BWQ-H does. 
Under the OU-based operation scheme, BSQ simply performs sequential computation on each OU and sequential activation data fetch from the IR. 
The indexing overhead for BSQ under the OU-based operation scheme is negligible. 
Nonetheless, the hardware performance of BSQ is lower than that of BWQ-H due to the fact that BWQ can achieve a higher model compression ratio with the fine-grained quantization granularity. 
On average, BWQ-H is able to achieve $1.45\times$ speedup and $2.66\times$ energy saving over BSQ. 

\subsection{Indexing Overhead Analysis}
\begin{figure}
    \centering
    \includegraphics[width=0.48\textwidth]{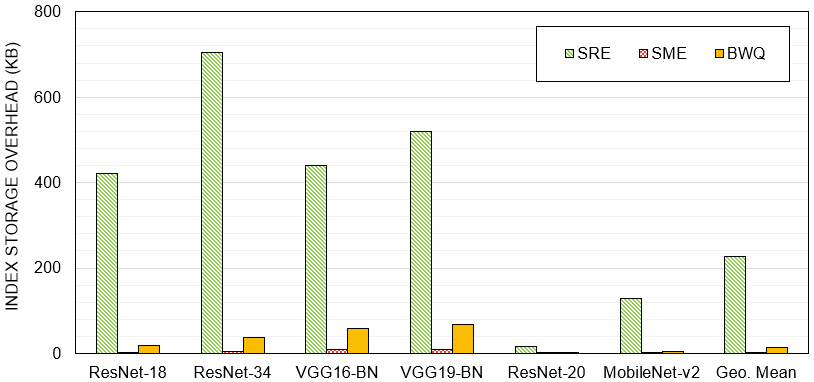}
    \caption{Indexing overhead comparison of CIFAR-10 models.}
    \label{idx-overhead}
\end{figure}

In this section, we evaluate the index overhead of BWQ-H, SRE, and SME when using the OU-based scheme for CIFAR-10 models which are mentioned in Table \ref{BWQ-A-results}. 
Here, $9 \times 8$ OUs are assumed. 
As shown in Fig. \ref{idx-overhead}, on average, BWQ-H's indexing overhead is $17.38 \times$ lower than SRE and $4.46 \times$ higher than SME.
As discussed in the previous section, as SRE exploits the extremely fine-grained OU-row level sparsity for both weight and activation, it suffers from significant index storage overhead.
SRE can require as much as 704KB to store the index for the ResNet-34 model. 
On the other hand, the memory consumption for storing the index in SME is much smaller, as it only explores the sparsity at the crossbar-row level. 
However, this coarse-grained compression granularity hinders SME from achieving the optimal model compression ratio.
As a result, BWQ-H can attain higher speedup and reduced energy consumption than SME.

\subsection{Ablation Studies}
\subsubsection{Regularization Strength and Re-quantization Interval}
\begin{figure}
    \flushright
    \includegraphics[width=0.48\textwidth]{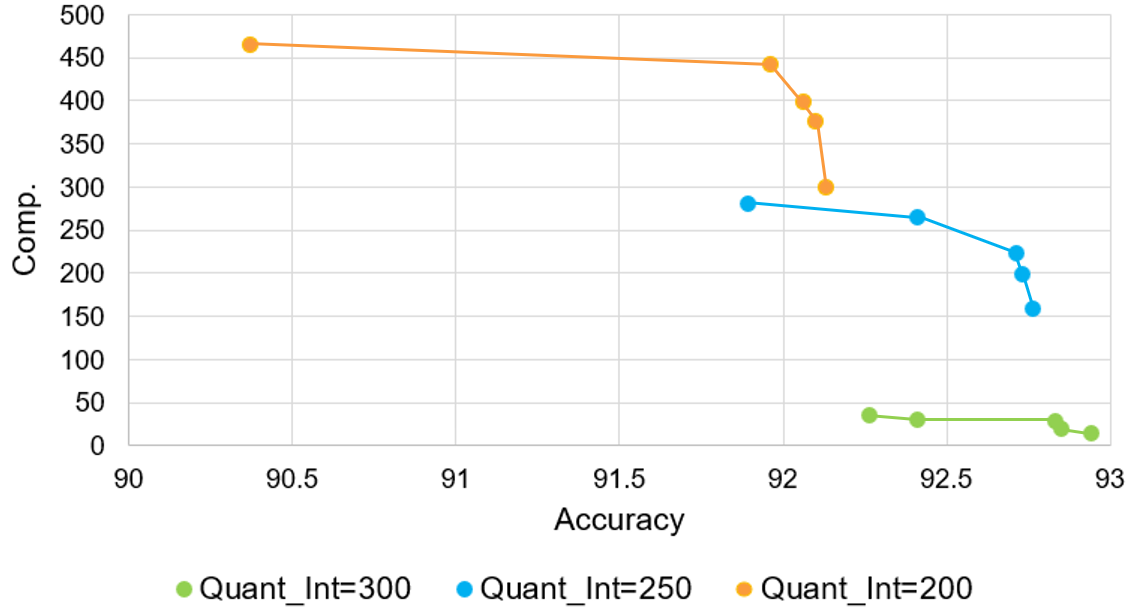}
    \caption{BWQ-A's accuracy and model compression ratio against varying regularization strength and re-quantization intervals.}
    \label{reg-strength}
\end{figure}

These are the two most important hyperparameters in BWQ-A that affect the model's accuracy and compression ratio. 
As shown in Fig. \ref{reg-strength}, we evaluate the performance of BWQ-A with VGG19-BN on the CIFAR-10 dataset.
Each data series represents 5 experiments, each with a different regularization strength: [5e-4, 1e-3, 3e-3, 5e-3, 1e-2]. 
It is understandable that applying larger regularization strength and more frequent re-quantization can enhance the model's compression ratio at the expense of accuracy.
It can be concluded from Fig. \ref{reg-strength} that the model's compression ratio is more influenced by the re-quantization interval than the regularization strength. 
It is evident from Fig. \ref{reg-strength} that using shorter re-quantization intervals consistently yields a better trade-off between compression ratio and accuracy. 
However, for VGG-19BN trained on CIFAR-10, using a re-quantization interval shorter than 200 introduces excessive quantization noise, making the training process unstable.

\subsubsection{OU Size}
\begin{figure}
    \flushright
    \includegraphics[width=0.48\textwidth]{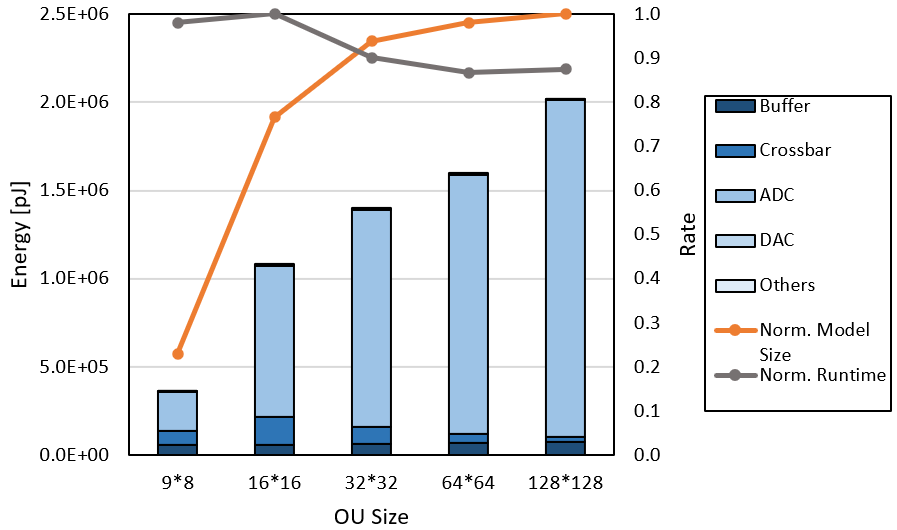}
    \caption{BWQ's energy consumption, normalized model size, and normalized runtime against varying OU sizes. 
    The bars correspond to the y-axis on the left, while
    the orange and black lines correspond to the y-axis on the right.}
    \label{varying_OU_size}
\end{figure}

In spite of the fact that the OU size for practical ReRAM-based accelerators is currently limited by the manufacturing technology of ReRAM devices, it is highly plausible that future advancements in manufacturing technology will enable the support of larger OU sizes. 
Therefore, we provide a future roadmap for practical ReRAM-based accelerators by examining the scalability of BWQ with varying OU sizes, from $9 \times 8$ to $128 \times 128$.

Fig. \ref{varying_OU_size} shows BWQ's performance with the ResNet-18 model tested on CIFAR-10 for varying OU sizes. Other models exhibit similar trends. 
As coarser quantization granularity results in a lower model compression ratio, the model size achieved with the BWQ-A quantization scheme increases with OU size. 
The model's inference runtime can be affected by a number of factors when OU size varies. 
When a larger OU is considered, then the computation with the entire crossbar can be completed within fewer clock cycles, thus reducing the runtime. However, a larger OU leads to a lower model compression ratio. Thus, more clock cycles are required to finish the computation with a larger model. 
As a larger OU contains more BLs, it also requires higher ADC precision, which leads to increased ADC latency. 
Overall, the runtime only shows minor variations with increasing OU size, reaching its minimum value at an OU size of 64 $\times$ 64. 
A similar analysis can be made for energy consumption. 
Fig. \ref{varying_OU_size} also shows the energy consumption breakdown of different components. 
ADC energy steadily increases with OU size since larger OUs require ADCs with higher precisions, and ADC energy scales up significantly with its precision. As ADC energy constitutes the majority of the energy consumption in our OU-based scheme, the overall energy consumption increases as the OU size grows. 
Therefore, if the primary goal is minimizing the runtime, a medium-sized OU would be preferable. On the other hand, if the task has a limited energy consumption budget, then the 9 $\times$ 8 OU configuration would be the most beneficial design option.


\section{Conclusion} \label{conclusion}
In this paper, we present BWQ, an algorithm-architecture co-design framework to enable highly efficient ReRAM-based DNN accelerators. 
Due to the practical concerns, BWQ adopts the OU-based operation scheme.
At the algorithm level, we introduce BWQ-A, a block-wise mixed precision quantization scheme. 
The size of the weight block aligns with the size of the OU. 
By employing finer quantization granularity, a high average model compression ratio of $58.27 \times$ is achieved within 1\% accuracy degradation compared to the floating-point baseline models. 
At the architecture level, we present the design of BWQ-H, which enables the efficient inference of the models quantized using BWQ-A. 
BWQ-H incorporates a novel precision-aware mapping scheme to increase memory utilization with an OU for the mixed-precision quantized models. 
Additionally, a memory controller is designed to generate control signals to the peripheral circuits to enable the computation of mixed-precision weights on the same crossbar. 
Leveraging the high model compression ratio, on average, BWQ-H achieves a $17.47 \times$ speedup and a $6.08 \times$ energy efficiency over existing ReRAM-based architectures.
We also analyze the scalability of BWQ-A and BWQ-H with varying OU sizes. 
This serves as a future road map for practical ReRAM-based accelerators depending on the evolution of the manufacturing technology of ReRAM devices.

\footnotesize
\bibliographystyle{ieeetr}
\bibliography{ref} 


\end{document}